\documentclass[manuscript]{aastex}
\begin{document}

\shorttitle{IPN Supplement}
\shortauthors{Hurley, K., Pal'shin, V., Briggs, M. et al.}

\title{The Interplanetary Network Supplement to the \it Fermi \bf GBM Catalog of Cosmic
Gamma-Ray Bursts }

\author{K. Hurley}
\affil{University of California, Berkeley, Space Sciences Laboratory,
7 Gauss Way, Berkeley, CA 94720-7450, USA}
\email{khurley@ssl.berkeley.edu}

\author{V. D. Pal'shin, R. L. Aptekar, S. V. Golenetskii, D. D. Frederiks, E. P. Mazets, D. S. Svinkin}
\affil{Ioffe Physical Technical Institute, St. Petersburg, 194021, Russian Federation}

\author{M. S. Briggs, V. Connaughton}
\affil{ University of Alabama in Huntsville, NSSTC, 320 Sparkman Drive, Huntsville, AL 35805, USA}

\author{C. Meegan}
\affil{Universities Space Research Association, NSSTC, 320 Sparkman Drive, Huntsville, AL 35805, USA}

\author{J. Goldsten}
\affil{Applied Physics Laboratory, Johns Hopkins University, Laurel, MD 20723, U.S.A.}

\author{W. Boynton, C. Fellows, K. Harshman}
\affil{University of Arizona, Department of Planetary Sciences, Tucson, Arizona 85721, U.S.A.}

\author{I. G. Mitrofanov, D. V. Golovin, A. S. Kozyrev, M. L. Litvak, A. B. Sanin}
\affil{Space Research Institute, 84/32, Profsoyuznaya, Moscow 117997, Russian Federation}

\author{A. Rau, A. von Kienlin, X. Zhang}
\affil{Max-Planck-Institut f\"{u}r extraterrestrische Physik,
Giessenbachstrasse, Postfach 1312, Garching, 85748 Germany}

\author{K. Yamaoka}
\affil{Division of Particle and Astrophysical Science, Graduate School of Science, Nagoya University,
Furo-cho, Chikusa-ku, Nagoya 464-8602, Japan}

\author{Y. Fukazawa, Y. Hanabata, M. Ohno}
\affil{Department of Physics, Hiroshima University, 1-3-1 Kagamiyama, Higashi-Hiroshima, Hiroshima 739-8526, Japan}

\author{T. Takahashi}
\affil{Institute of Space and Astronautical Science (ISAS/JAXA), 3-1-1 Yoshinodai, Sagamihara, Kanagawa 229-8510, Japan}

\author{M. Tashiro, Y. Terada}
\affil{Department of Physics, Saitama University, 255 Shimo-Okubo, Sakura-ku, Saitama-shi, Saitama 338-8570, Japan}

\author{T. Murakami}
\affil{Department of Physics, Kanazawa University, Kadoma-cho, Kanazawa, Ishikawa 920-1192, Japan}

\author{K. Makishima\altaffilmark{1}}
\affil{Department of Physics, University of Tokyo, 7-3-1 Hongo, Bunkyo-ku, Tokyo 113-0033, Japan}
\altaffiltext{1}{Makishima Cosmic Radiation Laboratory, The Institute of Physical and Chemical Research (RIKEN),
2-1 Hirosawa, Wako, Saitama 351-0198, Japan}

\author{S. Barthelmy, T. Cline\altaffilmark{2}, N. Gehrels} 
\affil{NASA Goddard Space Flight Center, Code 661, Greenbelt, MD 20771, U.S.A.}
\altaffiltext{2}{Emeritus}
\altaffiltext{3}{Joint Center for Astrophysics, University of Maryland, Baltimore County, 1000 Hilltop Circle, Baltimore, MD 21250}
\altaffiltext{4}{Universities Space Research Association, 10211 Wincopin Circle, Suite 500, Columbia, MD 21044}

\author{J. Cummings\altaffilmark{3}}
\affil{UMBC/CRESST/NASA Goddard Space Flight Center, Code 661, Greenbelt, MD 20771, U.S.A.}

\author{H. A. Krimm\altaffilmark{4}}
\affil{CRESST/NASA Goddard Space Flight Center, Code 661, Greenbelt, MD 20771, U.S.A.}

\author{D.M. Smith}
\affil{Physics Department and Santa Cruz Institute for Particle Physics,
University of California, Santa Cruz, Santa Cruz, CA 95064, U.S.A.}

\author{E. Del Monte, M. Feroci}
\affil{INAF/IASF-Roma, via Fosso del Cavaliere 100, 00133, Roma, Italy}

\author{M. Marisaldi}
\affil{INAF/IASF-Bologna, Via Gobetti 101, I-40129 Bologna, Italy}

\begin{abstract}

We present Interplanetary Network (IPN) data for the gamma-ray bursts in the first
\it Fermi \rm Gamma-Ray Burst Monitor (GBM) catalog.  Of the 491 bursts in that
catalog, covering 2008 
July 12 to 2010 July 11, 427 were observed by at least one other instrument in the 9-spacecraft IPN.
Of the 427, the localizations of 149 could be improved by arrival time analysis
(or ``triangulation'').  For any given
burst observed by the GBM and one other distant spacecraft, triangulation gives an annulus of possible arrival
directions whose half-width varies between about 0.4$\arcmin$ and 32$\arcdeg$, depending
on the intensity, time history, and arrival direction of the burst, 
as well as the distance between the spacecraft.  We find that the IPN localizations
intersect the 1$\sigma$ GBM error circles in only 52\% of the cases, if no systematic
uncertainty is assumed for the latter.  If a 6$\arcdeg$ systematic uncertainty
is assumed and added in quadrature, the two localization samples agree about 87\% of the
time, as would be expected.  If we then multiply the resulting error radii by a factor
of 3, the two samples agree in slightly over 98\% of the cases, providing a good estimate
of the GBM 3$\sigma$ error radius.   The IPN 3$\sigma$ error boxes
have areas between about 1 square arcminute and 110 square degrees, and are, on the average,
a factor of 180 smaller than the corresponding GBM localizations. 
We identify two bursts in the IPN/GBM sample that did not appear in the GBM catalog.  In
one case, the GBM triggered on a terrestrial gamma flash, and in the other, its
origin was given as ``uncertain''.  We also discuss the sensitivity and calibration of the
IPN.

\end{abstract}


\section{Introduction}

This paper presents the latest in a series of catalogs of gamma-ray burst (GRB) localizations 
obtained by arrival time analysis, or ``triangulation'' between the spacecraft in the 3rd interplanetary network (IPN)
(table~\ref{ipncatalogs}).  In the present paper, we present the localization data on 149 bursts which occurred 
during the period covered by the first, two-year \it Fermi \rm GBM GRB catalog \citep{p1}
(2008 July 12 to 2010 July 11).  As the composition of the IPN has changed over the years, we
present a summary of the instrumentation and techniques in the following section.  Section 3 contains the localization data,
which are also available on the IPN website\footnote[5]
{http://ssl.berkeley.edu/ipn3/index.html}.  In section
4, we discuss the statistics of the localizations.

\section{Technique, Instrumentation, Calibration, and Sensitivity}

The triangulation technique is illustrated in figure 1.
When a GRB arrives at two spacecraft with a delay $\rm \delta$T, it may be
localized to an annulus whose half-angle $\rm \theta$ with respect to the
vector joining the two spacecraft is given by 
\begin{equation}
cos \theta=\frac{c \delta T}{D}
\end{equation}
where c is the speed of light and D is the distance between the two
spacecraft.  (This assumes that the burst is a plane wave, i.e. that its
distance is much greater than D.)  The annulus width d$\rm \theta$, and thus one dimension of
the resulting error box, is 
\begin{equation}
d \rm \theta =c \rm \sigma(\delta T)/Dsin\rm \theta
\end{equation}
where
$\rm \sigma(\delta$T) is the uncertainty in the time delay.  
The radius of each annulus and the right ascension and declination 
of its center are calculated in a heliocentric (i.e., aberration-corrected) frame.

The composition of the missions and experiments comprising the interplanetary network changes as old missions are terminated and
new missions are introduced. During the period covered in the present catalog, the IPN consisted of
\it Konus-Wind \rm, at distances up to around 5 light-seconds from Earth \citep{a1}; \it Mars Odyssey\rm, in orbit around Mars at
up to 1250 light-seconds from Earth \citep{h13}; \rm the \it International Gamma-Ray
Laboratory \rm (INTEGRAL), in an eccentric Earth orbit at up to 0.5 light-seconds from Earth \citep{r1}; \rm the \it Mercury Surface, Space Environment, Geochemistry, and Ranging
\rm mission (MESSENGER), launched in 2004 August, and in an eccentric orbit around Mercury beginning 2011 March 18, up to 690 light-
seconds from Earth \citep{g1}; 
and \rm the \it Ramaty High Energy Solar Spectroscopic Imager \rm (RHESSI) \citep{s1}, \it Swift \rm \citep{g2}, \it Fermi \rm \citep{m2}, \it Suzaku \rm \citep{t1,y1}, and AGILE \citep{m1,d1,t2}, all in low Earth orbit. {

The detectors in the IPN vary widely in shape, composition, time resolution, and energy range.  Also, onboard timekeeping techniques 
and accuracies are
not the same from mission to mission, and spacecraft ephemeris data are given only as predicts for some missions.  Since the
accuracy of the triangulation technique depends on all these parameters, end-to-end calibrations and sensitivity checks are
a constant necessity.  For the current IPN, we utilize the following method.  For every burst for which the \it Swift \rm 
X-Ray Telescope (XRT) detects an X-ray afterglow, we search for GRB detections in all the IPN experiments.  If the burst
was detected by a) \it Odyssey \rm and \it Konus \rm or by \it Odyssey \rm and a near-Earth mission, b) MESSENGER and
\it Konus \rm or by MESSENGER and a near-Earth mission, or c) \it Konus \rm and a near-Earth mission, we derive
an IPN annulus by triangulation.  We then calculate the angle between the annulus center line and the XRT position $\theta_X$, 
taken from the GCN Circulars, and which
we take to be a point source, because its positional uncertainty is much less than the annulus width $d\theta$ (figure 2).  $d\theta$ is calculated
such that the distribution of annulus widths is approximately Gaussian, so the distribution of $\theta_X$/$d\theta$ should follow
a normal distribution with mean zero and standard deviation 1, if systematic uncertainties are neglected.  We have used this procedure
so far for 78 MESSENGER/Konus or MESSENGER/near-Earth triangulations, 292 \it Konus\rm/near-Earth triangulations, and 72 
\it Odyssey/Konus \rm or \it Odyssey\rm/near-Earth triangulations.  We find that for the interplanetary spacecraft a systematic uncertainty equal to roughly 0.75 times
the statistical one is required to make the distributions consistent with normal distributions.  An example is shown in figure 3.
Systematic uncertainties arise from numerous sources.  Some are certainly negligible in some cases, while others may be important.  But in almost all cases, 
it is impossible to assign an accurate number to them.  A partial list follows, in no particular order.

\begin{enumerate}

\item Variations in the clock accuracy from one spacecraft to another.  Different spacecraft have different ways of calibrating their clocks, and assigning times to the 
time bins of GRB time histories.  We know, for example, that in some cases, the GRB timing is subject to uncertainties, even though the spacecraft oscillator is quite accurate.
\item Predict timing.  In many cases, the time assigned to a GRB is a predicted time, and it is never updated.  In other cases, such as \it Odyssey \rm and MESSENGER, the time is eventually updated
using an accurate model for the clock drift; in this study, the updated times have been used for these spacecraft.  In other cases, no final clock corrections are applied.
\item Different time resolutions.  For any given spacecraft pair, the time resolutions can be vastly different, and sometimes one is not an exact multiple of the other.  One time history is adjusted to match the time resolution of the other spacecraft in the light curve comparisons.  This can be done in a variety of ways, but each is subject to uncertainties.  Even in cases where one time resolution is in principle an exact multiple of another, the true values of the bin widths can be slightly different from their nominal values due to different on-board electronics.
\item Spacecraft ephemerides.  Some ephemerides are predicts, while others are final.  In these comparisons, the final ephemerides were used where possible, but they were not always available.  
\item Different energy responses of the various detectors.  In most cases, the GRB light curves are recorded in different energy ranges from one another.  Even in those cases where we attempt to match the energy ranges of  the detectors (i.e. where the photons are energy-tagged), the detector responses within those ranges are different due to the very different detector designs.

\end{enumerate}

It is often possible to derive very precise triangulation annuli for bursts detected by \it Konus \rm and the GBM, even though the distance between the spacecraft is not large.
The reasons are first, that the first two seconds of triggered \it Konus \rm data are transmitted with 2 ms time resolution; this is the finest resolution of all the IPN detectors which bin their data.
Second, GBM time- and energy-tagged data can be utilized to match Konus' time bins and energy range, minimizing two possible sources of systematic uncertainties.  Thus for short-duration or
intense GRBs, or bursts with fine time structure, \it Konus \rm/GBM annulus widths as small as several arcminutes can be obtained \citep{p2}. 
To verify Konus-GBM triangulations we have derived Konus-GBM triangulation annuli for 52 precisely localized bursts. The 3$\sigma$ half-widths of these annuli range from 0$\fdg$11 to 21$\fdg$8 with a mean of 3$\fdg$0 and a geometrical mean of 1$\fdg$19. Figure 4 shows the distribution of the offsets (in sigma) of the precise positions. The mean offset is 0.09 and the standard deviation is 0.77. The minimum offset is -1.40 and the maximum is 1.69.

To calibrate the IPN sensitivity, we use estimates of the peak fluxes, fluences, durations, and E$_{peak}$ of a large number of GBM bursts\footnote[6]
{http://heasarc.gsfc.nasa.gov/cgi-bin/W3Browse/w3query.pl}.
These are measured in the 50 -- 300 keV range, and are given in $\rm photons \, cm^{-2} \, s^{-1}$ (measured over a 1024 ms period), $\rm erg \, cm^{-2} $,
seconds, and keV, respectively.  At the time this catalog was submitted in its final version, there were 1078 GBM bursts with peak flux, fluence, and duration
entries, and 482 with E$_{peak}$ entries.  To calculate the IPN sensitivity, we determined a) whether any other
IPN spacecraft also detected the burst, and b) whether 
Konus, MESSENGER, or \it Odyssey \rm detected the burst.  Only the latter detections can lead to meaningful triangulations, because of
their larger distances from Earth.  From these numbers, we calculate the detection probabilities as functions of flux, fluence, duration, and E$_{peak}$.
The results are shown in figures 5, 6, 7, and 8.  In each of the four graphs, the detection probabilities (or IPN efficiencies) represent
an integral over the other 3 variables, as well as over duty cycles, and, for all the instruments
except Konus, planet-blocking.  The probabilities of IPN detections are 50\% or greater for peak fluxes in the range $\rm 1 - 3 \, photons \, cm^{-2} \, s^{-1}$
and for fluences in the range $\rm 1 - 3 \times 10^{-6}\, erg \, cm^{-2}$.

Every cosmic burst detected by the GBM was searched for in the
IPN data; GBM localizations were used to calculate arrival time windows for \it Odyssey \rm and
MESSENGER, but the total crossing time windows defined by light-travel times were
examined in all cases. The resulting detections are given in table~\ref{detections}.  (Note that this
table supersedes the information in table 2 of \citet{p1}, which is incomplete.)  \it Konus \rm and \it 
Suzaku \rm can detect bursts in both a triggered (2 -- 64 ms time resolution) and an untriggered (1 -- 3 s time
resolution) mode; both modes are counted as detections in this table.  Also, detections by several instruments
which are not part of the IPN have been reported in the table, namely the \it Fermi \rm LAT, 
MAXI, (\it Monitor of All-sky X-ray Image\rm), and RXTE (\it Rossi X-Ray Timing Explorer\rm).

Two events in table~\ref{detections} were detected by the GBM, but did not appear in the GBM catalog.
The origin of GRB 091013 was classified as ``uncertain''.  However, the cosmic nature of this
event is confirmed by \it Konus\rm.  GRB 100501 was detected by numerous IPN spacecraft, including
the GBM.  However, in that case, the actual GBM trigger was caused by a terrestrial gamma flash.

Whenever \it Konus \rm (in triggered, high time resolution mode), \it Odyssey, \rm or MESSENGER detected the
burst, we calculated one or more triangulation annuli. 
The annuli are given in table~\ref{annuli}, and an example is shown in figure 9.  In general, the annuli obtained by triangulations are small circles on the celestial
sphere, so their curvature, even across a relatively small GBM error circle, may not be negligible, so that a simple, four-corner error box cannot always be defined accurately. 
For this reason, we do not cite the intersection points of
the annuli with the error circles.  A prescription for deriving these points, however, may be found in \citet{h3}. 

When three widely separated experiments observe a burst, the result is two annuli which generally 
intersect to define two small error boxes.  The proximity to the GBM error circle may be used to distinguish the correct
one. An example is shown in figure 10.  When \it Konus, Odyssey, \rm MESSENGER and a near-Earth
spacecraft (including INTEGRAL) detect a burst, the position is over-determined.  In these
cases, a goodness-of-fit can be derived for the localization, and an error ellipse can be generated \citep{h5}. 
Although we utilize this procedure whenever possible, we do not quote the localizations as
error ellipses in this catalog, because, like the annuli, their curvature can render a simple
parameterization inaccurate.  A number of degenerate cases can occur in a three-spacecraft
triangulation; they are discussed in \citet{h11}.

When \it Konus \rm and an interplanetary or near-Earth spacecraft observe a burst, it is often possible to define
a long, narrow error box from \it Konus' \rm determination of the burst's ecliptic latitude (figure 11).
This is derived from a comparison of the count rates on the two \it Konus \rm detectors, and
its accuracy is generally of the order of $\pm \rm 10 \arcdeg$.  A study of over
1800 \it Konus \rm events indicates that the ecliptic latitude limits determined in this way
can be considered to be an $\sim \rm 95\%$ confidence band.  Systematic uncertainties usually
prevent a more accurate determination.

IPN annulus widths are often comparable to, or smaller than, \it Fermi \rm LAT error circle
radii, and can therefore reduce the areas of LAT localizations.  Figure 12 shows the example
of GRB 090323 \citep{o1,h14}, for which a \it Swift \rm ToO observation led to the discovery of an XRT 
\citep{k2}, optical \citep{u1}, and radio \citep{h15} counterpart.

\section{Table of IPN Localizations}

The 21 columns in table~\ref{annuli} give:
1) the date of the burst, in yymmdd format; this contains a link to a figure
on the IPN website showing the annulus or error box and the GBM error circle, 
2) the Universal Time of the burst at Earth,
3) the GBM right ascension of the center of the error circle (J2000), in degrees,
4) the GBM declination of the center of the error circle (J2000), in degrees,
5) the 1$\sigma$ statistical GBM error circle radius, in degrees,
6) the right ascension of the center of the first IPN 
annulus, epoch J2000, in the heliocentric frame, in degrees 
7) the declination of the center of the first IPN 
annulus, epoch J2000, in the heliocentric frame, in degrees,
8) the angular radius of the first IPN annulus, in the heliocentric
frame, in degrees, 
9) the half-width of the first IPN annulus, in degrees; the 3$\sigma$
confidence annulus is given by R$_{\rm IPN1}$ $\pm$ $\delta$ R$_{IPN1}$, 
10) the right ascension of the center of the second IPN 
annulus, epoch J2000, in the heliocentric frame, in degrees,
11) the declination of the center of the second IPN 
annulus, epoch J2000, in the heliocentric frame, in degrees,
12) the angular radius of the second IPN annulus, in the heliocentric
frame, in degrees, 
13) the half-width of the second IPN annulus, in degrees; the 3$\sigma$
confidence annulus is given by R$_{\rm IPN2}$ $\pm$ $\delta$ R$_{IPN2}$,
14 and 15) the Konus ecliptic latitude band, in degrees,
16, 17, and 18) the right ascension, declination, and angular radius of
the Earth or Mars, if the planet blocks part of the localization, in degrees, and
19, 20, and 21) any other localization information, in right ascension, declination,
and angular radius, in degrees.  

The GBM data have been taken from the HEASARC online catalog\footnote[7]
{http://heasarc.gsfc.nasa.gov/cgi-bin/W3Browse/w3query.pl}, if the 
localization source was ``Fermi, GBM''.  For bursts with other localization
sources, the ``human-in-the-loop'' localization was used (V. Connaughton, private
communication, 2012).  GBM localizations are subject to change, and are
given here for convenience only.  The latest online
catalog should be considered to be the most authoritative source of the
up-to-date GBM data.   The data in table~\ref{annuli} are also available electronically\footnote[8] 
{http://ssl.berkeley.edu/ipn3/index.html}.  

\section{A Few Statistics}

There are 491 bursts in the GBM catalog \citep{p1}.  Of these, 427 (87\%) were observed by at least
one other IPN spacecraft.  They are listed in table~\ref{detections}, and the number of bursts observed
by each IPN spacecraft is compiled in table~\ref{ipnspacecraft}.  Those events which were not
observed by an IPN spacecraft had fluences between $\rm 4.5 \times 10^{-8} \, and \, 9.5 \times 10^{-6}
erg \, cm^{-2}$, peak fluxes between $\rm 0.33 \, and \, 8.8 \, photons \, cm^{-2} \, s^{-1}$, and durations between 0.13
and 218 s, as measured by the GBM \citep{g2,p1}.  
For 149 of them, it
was possible to improve the localizations by triangulation.  
The minimum and maximum 3$\sigma$ IPN annulus half-widths were $7.40 \times 10^{-3} \,\rm and \, 31.9 \arcdeg $, and the
average was 1.8$\arcdeg$.   The IPN error boxes have 3$\sigma$ areas between about 1 square arcminute and 110 square degrees.
Each IPN localization was compared to its corresponding GBM error
circle, as given in the online catalog\footnote[9]{http://heasarc.gsfc.nasa.gov/cgi-bin/W3Browse/w3query.pl}.
In that catalog, the GBM localizations have been approximated as circles, with 1$\sigma$ (statistical
only) radii.  Assuming that they are described by a two-dimensional normal distribution, we
would expect 87\% of the 3$\sigma$ IPN localizations to agree with them (i.e. to have some
intersection with them).  We find only 52\% agreement.    

If a GBM systematic uncertainty of 6$\arcdeg$ 
is assumed, and added in quadrature to the statistical uncertainty, we find the expected
87\% agreement.  If that radius is then multiplied by three, the agreement becomes 98\% (3
events with discrepant localizations), so
that this can be taken as an approximation to a 3$\sigma$ GBM confidence region for this particular
GRB sample.  A more detailed
analysis of systematics is given in \citet{c1}.  Comparing each IPN area with its corresponding
3$\sigma$ GBM area, as approximated above, we find an average reduction in area of a factor of
180.
 
\section{Discussion and Conclusion}

The \it Fermi \rm GBM has proven to be a worthy successor to BATSE.  It detects
about 245 GRBs per year and distributes their coordinates almost instantaneously
to a wide astronomical community.  The nine-spacecraft IPN is a good complement
to it, just as it was to BATSE.  It detects a total of about 325 bursts per year (18 per year are short-
duration, hard spectrum GRBs; see \citet{p2}), has virtually no planet-blocking
or duty cycle restrictions when all the spacecraft are considered, and it is capable
of good localization accuracy at the cost of longer delays.  There are many ground-based
experiments, both electromagnetic and non-electromagnetic, which can take advantage of the 
smaller IPN error boxes, and for which delays are not an issue.  In this sense, the GBM
and the IPN both expand the reach of \it Swift\rm, by localizing bursts which \it Swift \rm cannot.
For example, a search for gravitational
radiation is in progress which utilizes the IPN data on over 500 GRBs, the most extensive such search
to date; another search has begun for neutrinos, using IceCube data and almost 1000 IPN events.  

This catalog represents the first installment of the IPN supplements to the GBM burst catalogs.
Work is proceeding on the localization of IPN bursts observed during the 3rd and 4th
years of GBM operation. Data on some of these events may be found at
the IPN web site\footnote[10]{http://ssl.berkeley.edu/ipn3/interpla.html}.

\section{Acknowledgments}

Support for the IPN was provided by NASA grants NNX09AU03G (\it Fermi\rm), NNX08AC90G and NNX08AX95G
(INTEGRAL), NNX08AN23G and NNX09AO97G (\it Swift\rm), NNX08AZ85G and NNX09AV61G (\it Suzaku\rm), NNX07AR71G (MESSENGER),
and JPL Contracts 1282043 and Y503559 (\it Odyssey\rm).  The Konus-Wind experiment is supported by a
Russian Space Agency contract and RFBR grant 12-02-00032-a. This research has made use of data and/or software 
provided by the High Energy Astrophysics Science Archive Research Center (HEASARC), which is a service of the 
Astrophysics Science Division at NASA/GSFC and the High Energy Astrophysics Division of the Smithsonian Astrophysical Observatory.

\clearpage

\begin{figure}
\plotone{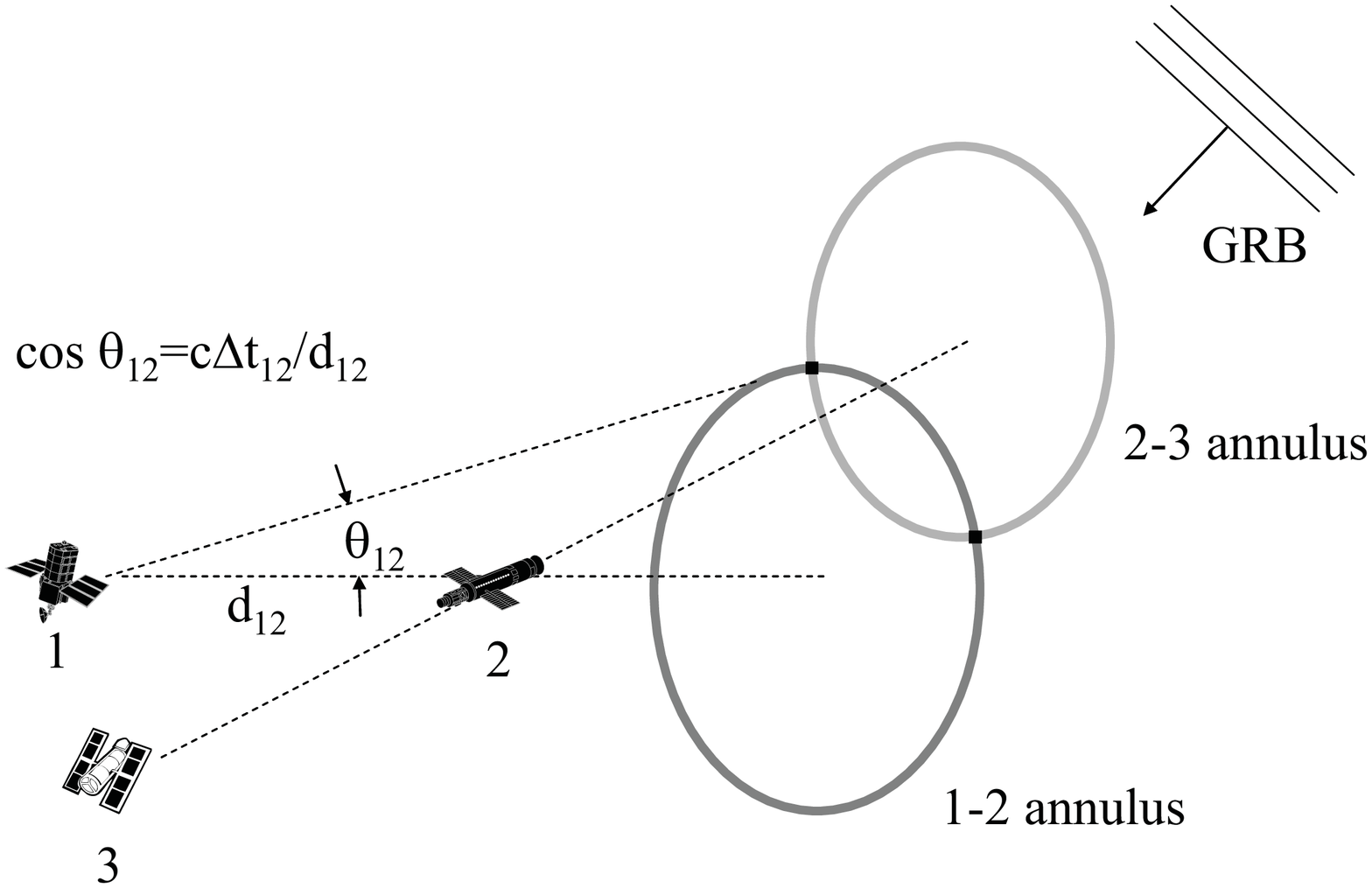}
\caption{The triangulation technique.  Each independent spacecraft pair is used
to derive an annulus of location for the burst.  Three spacecraft produce two
possible error boxes.  The ambiguity can be eliminated by the addition of a fourth,
non-coplanar spacecraft, by the anisotropic response of one of the experiments,
or by the GBM localization.}
\end{figure}

\begin{figure}
\plotone{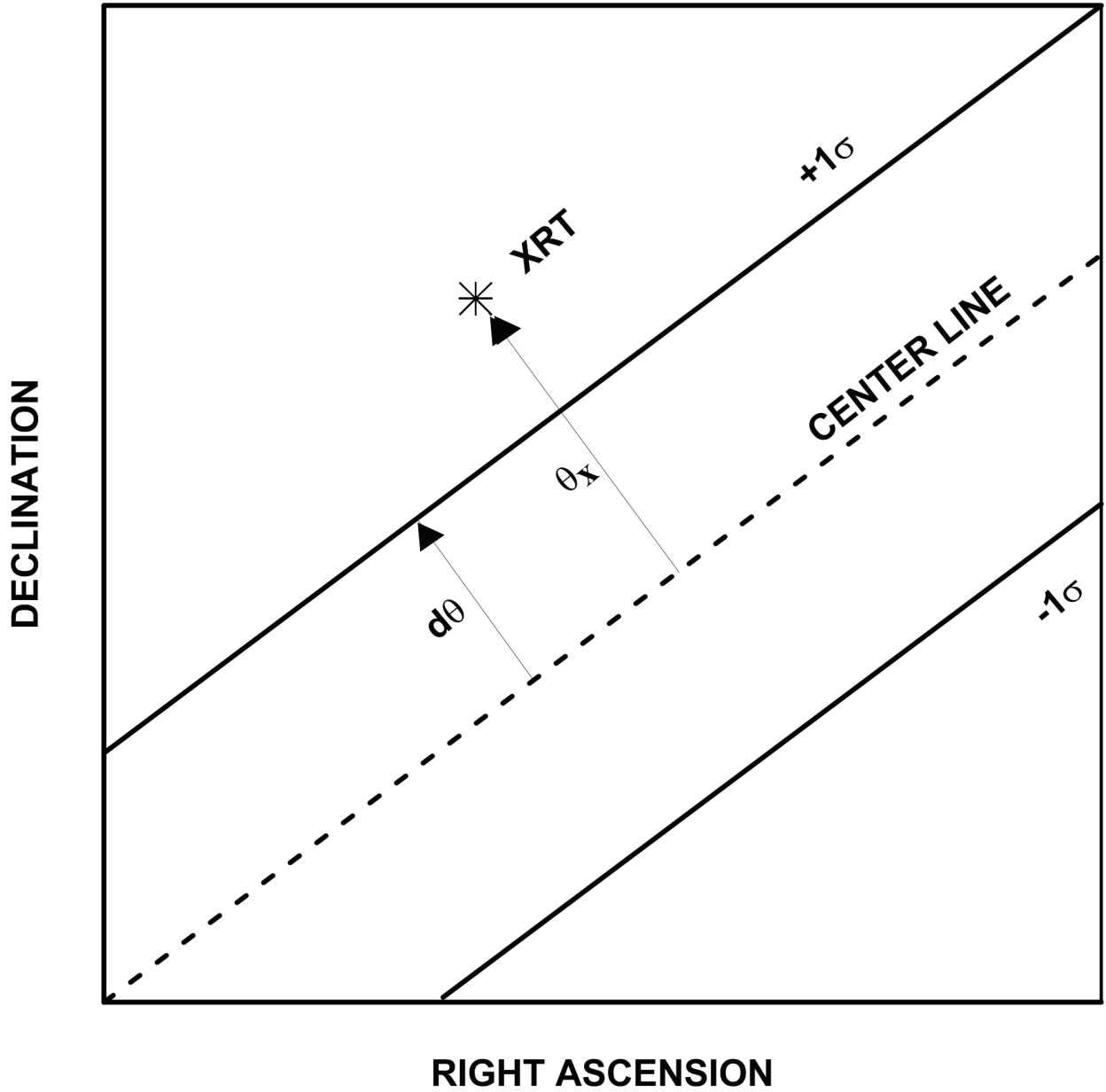}
\caption{Closeup of a portion of a triangulation annulus.  The dashed line
is the center line, and the two solid lines are the 1 $\sigma$ contours.  The 1 $\sigma$
annulus width is $d \theta$, and the minimum angle between the center line and
the XRT counterpart is $\theta_X$.}
\end{figure}

\begin{figure}
\plotone{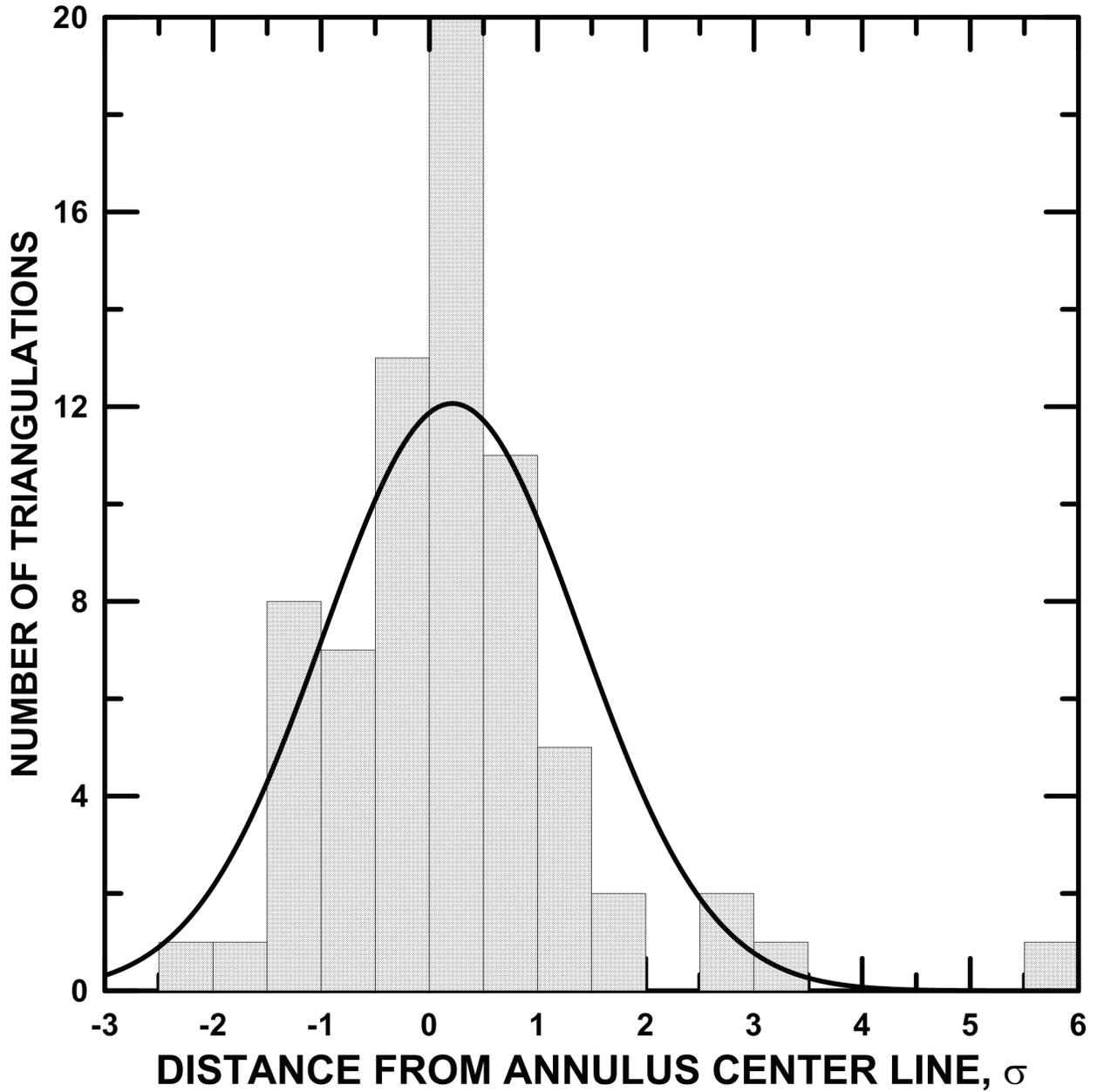}
\caption{MESSENGER triangulation accuracy.  The histogram shows the angles between
the annuli center lines and the XRT counterparts for 78 bursts.  The mean is 0.11 and
the standard deviation is 0.96.  A systematic uncertainty equal to .75 times the statistical uncertainty has
been assumed.  The solid line is
a Gaussian fit to the histogram.  }
\end{figure}

\begin{figure}\plotone{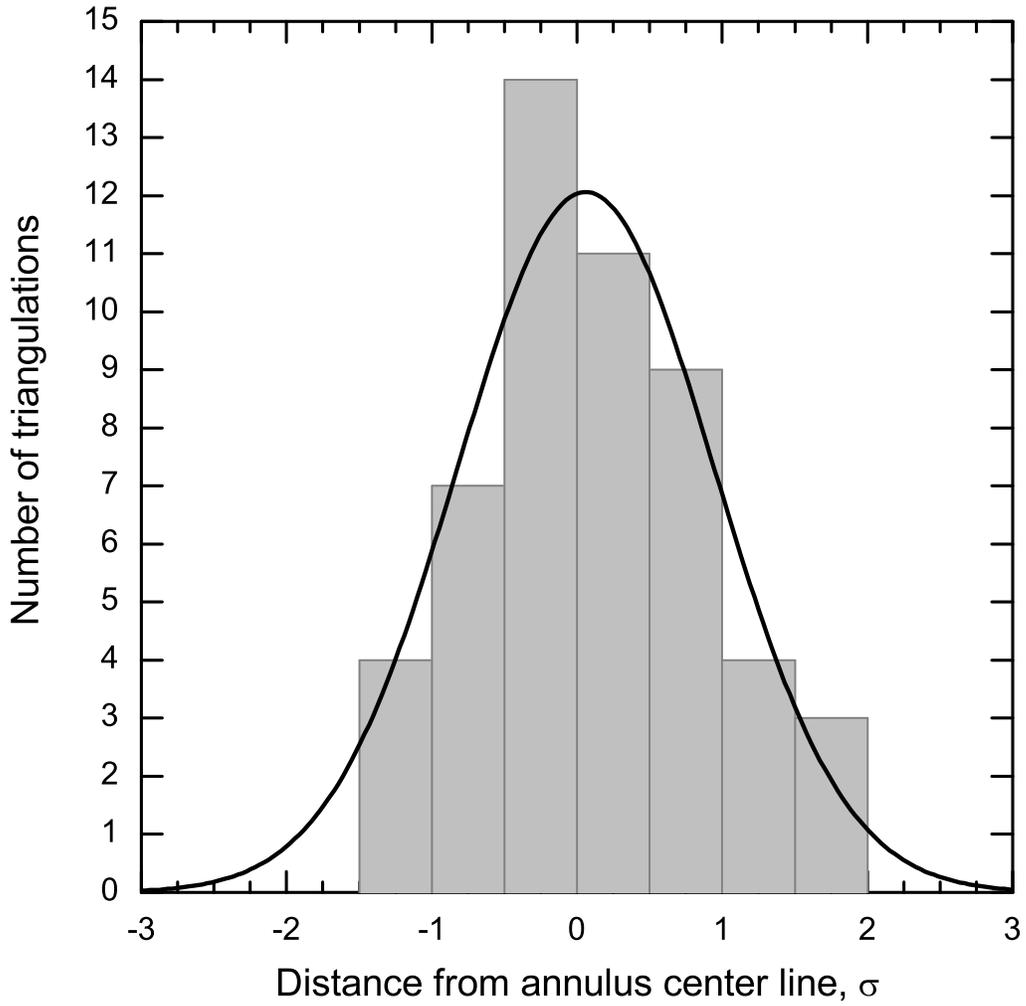}
\caption{Accuracy of Konus-GBM triangulations. The histogram shows the relative offsets in sigma between the annuli center lines 
and the XRT counterparts for 52 bursts. The mean is 0.09 and the standard deviation is 0.77. The solid line is a Gaussian fit to the histogram.} 
\end{figure}

\begin{figure}
\plotone{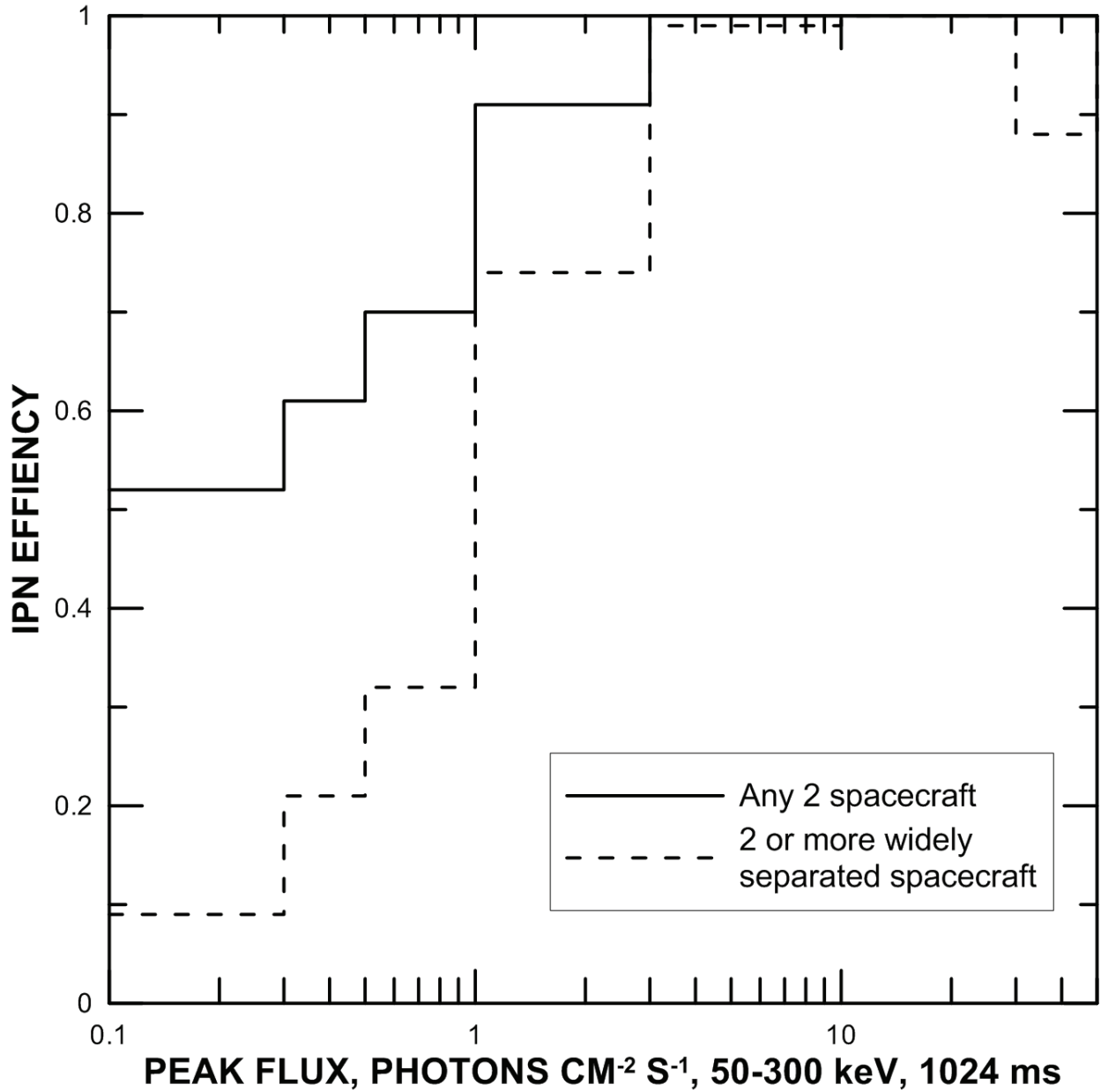}
\caption{The IPN efficiency as a function of GRB peak flux.  The peak flux is measured
over a 1024 ms time interval by the GBM in the 50 -- 300 keV energy range.  Two efficiencies
are shown.  The solid line is the probability that any IPN experiment (other than the GBM)
will detect the burst.  The dashed line is the probability that Konus, \it Odyssey, \rm 
or MESSENGER will detect it. Only the latter detections lead to accurate
triangulations. }
\end{figure}

\begin{figure}
\plotone{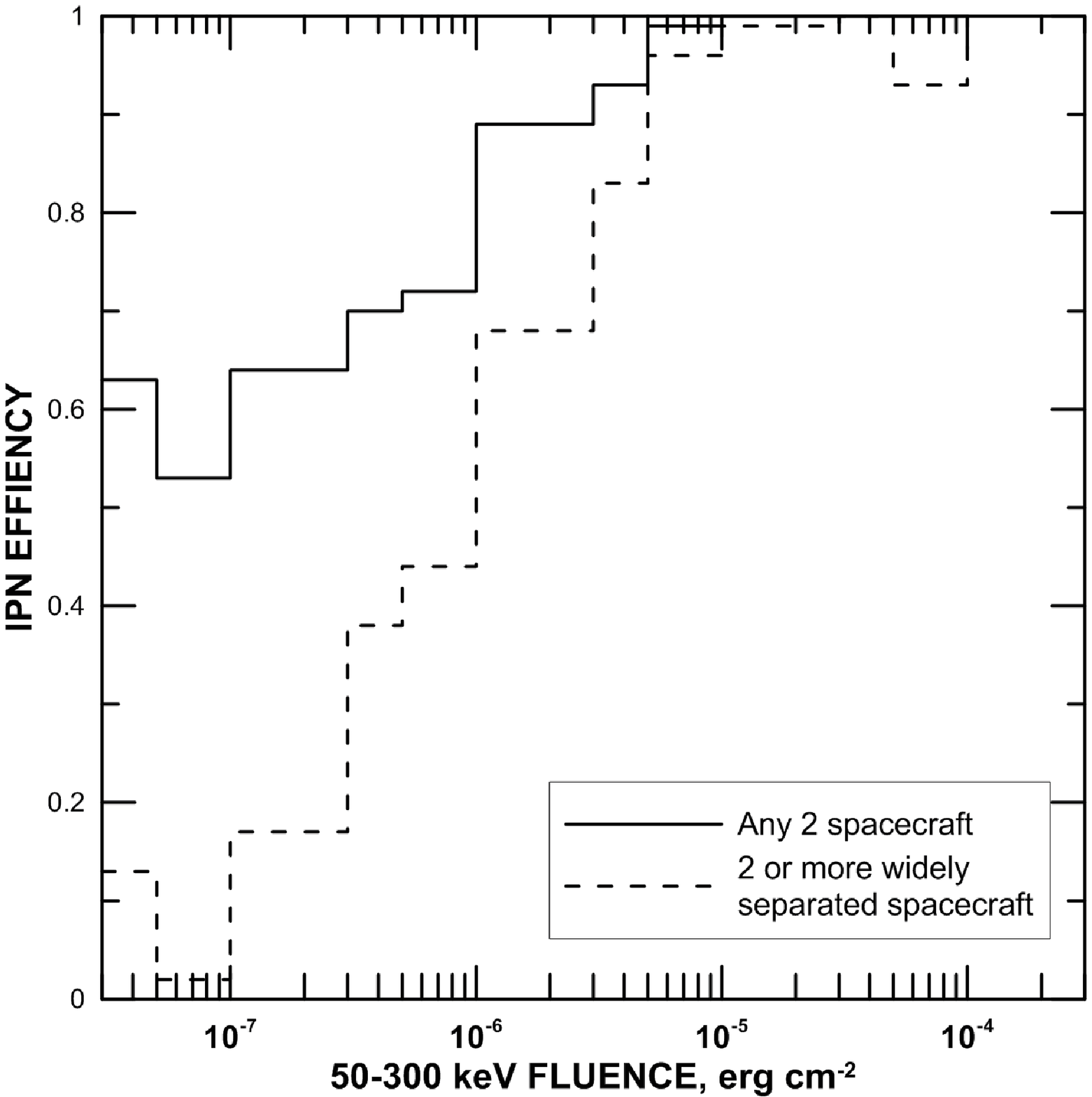}
\caption{The IPN efficiency as a function of GRB fluence.  The fluence is measured
by the GBM in the 50 -- 300 keV energy range.  Two efficiencies
are shown.  The solid line is the probability that any IPN experiment (other than the GBM)
will detect the burst.  The cashed line is the probability that Konus, \it Odyssey, \rm 
or MESSENGER will detect it.  Only the latter detections lead to accurate
triangulations.}
\end{figure}

\begin{figure}
\plotone{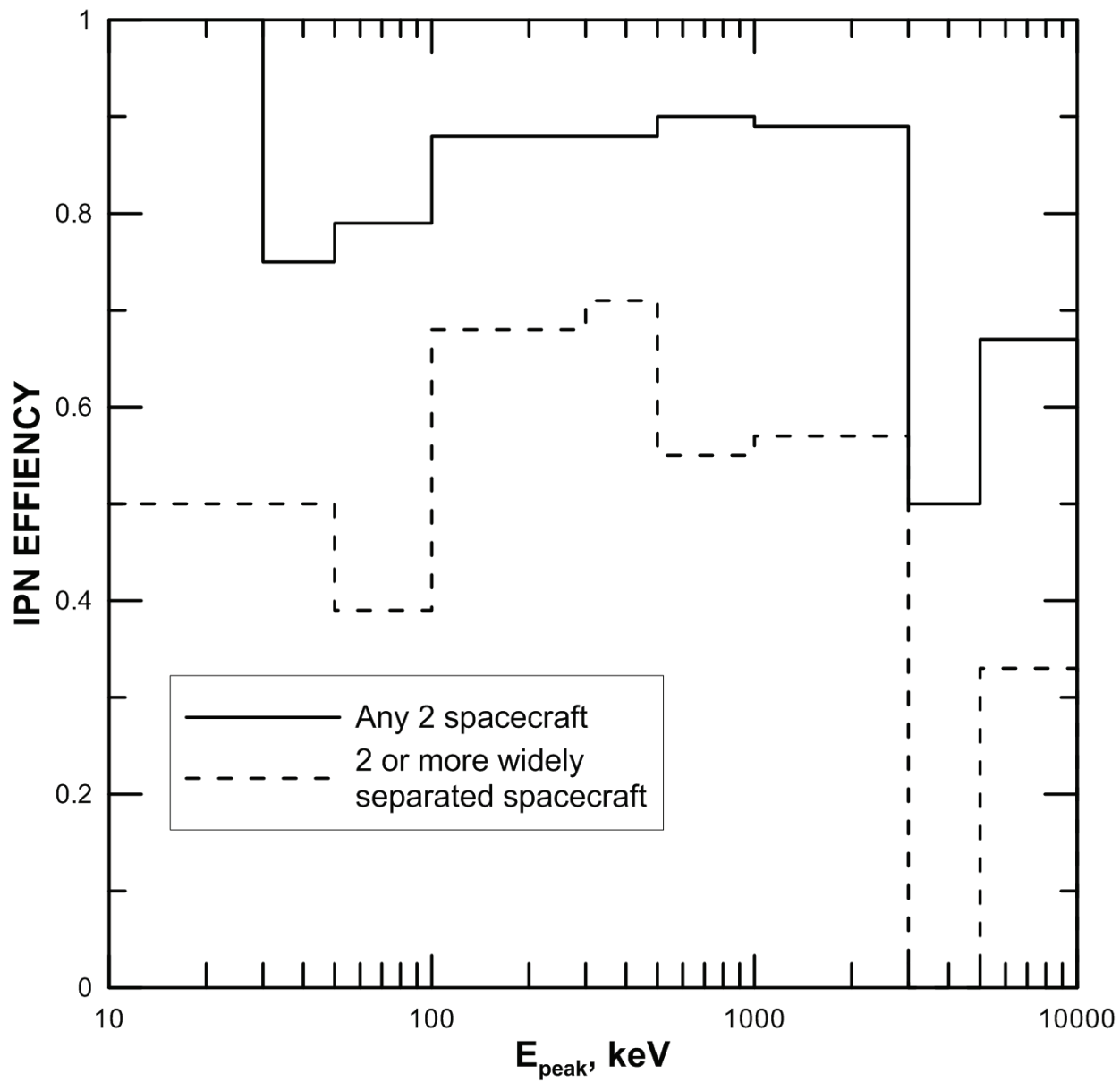}
\caption{The IPN efficiency as a function of GRB E$_{peak}$.  As measured by the GBM, this
is from a Band function fit to a single spectrum over the time range of the peak flux of the burst.  
Two efficiencies
are shown.  The solid line is the probability that any IPN experiment (other than the GBM)
will detect the burst.  The dashed line is the probability that Konus, \it Odyssey, \rm 
or MESSENGER will detect it.  Only the latter detections lead to accurate
triangulations.  The first and last two bins are based on 12 or fewer events, and have poor statistics.}
\end{figure}

\begin{figure}
\plotone{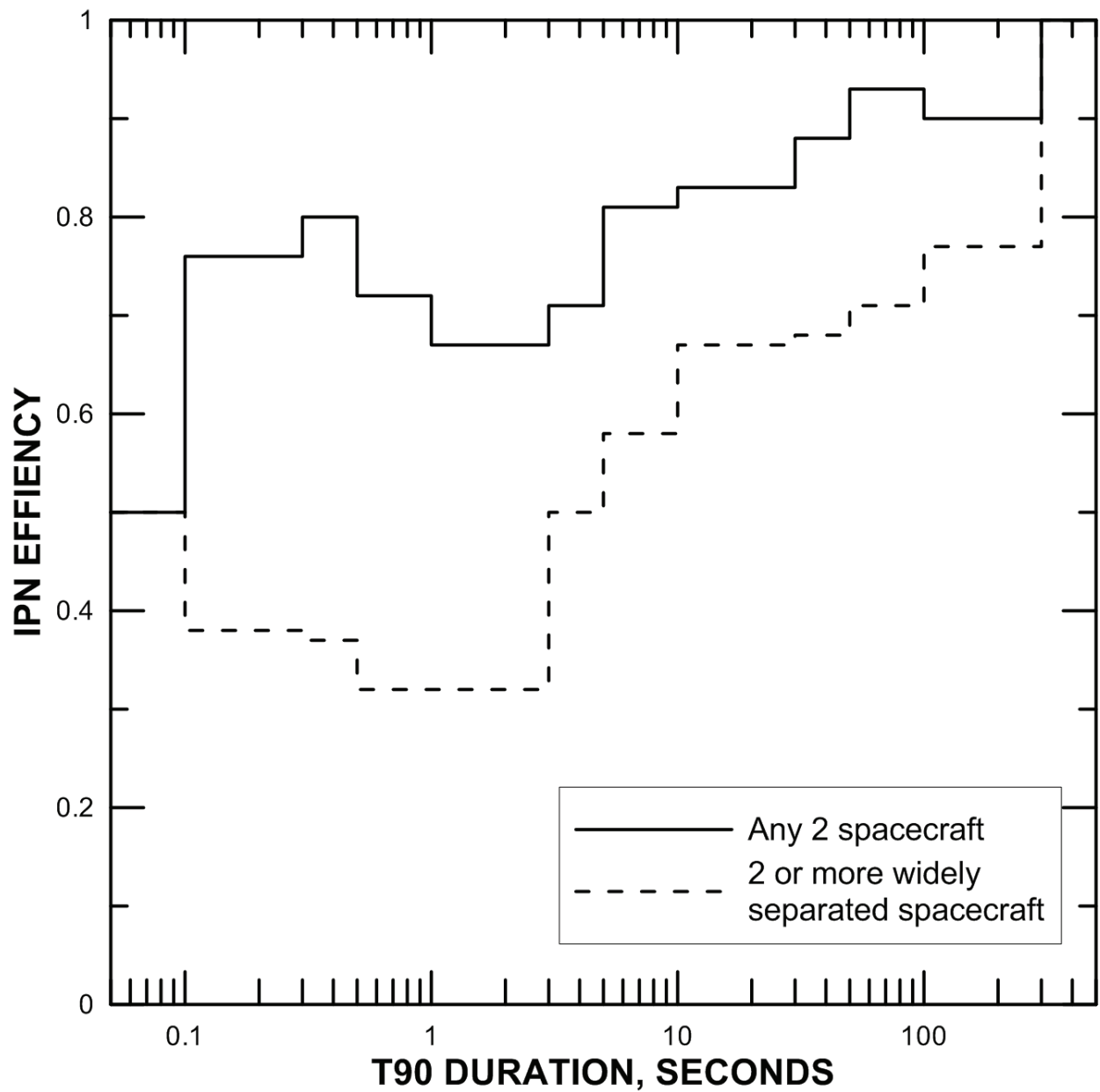}
\caption{The IPN efficiency as a function of GRB duration.  As measured by the GBM, this
is T90 in the 50 -- 300 keV energy range.  
Two efficiencies
are shown.  The solid line is the probability that any IPN experiment (other than the GBM)
will detect the burst.  The dashed line is the probability that Konus, \it Odyssey, \rm 
or MESSENGER will detect it.  Only the latter detections lead to accurate
triangulations.  The first and last two bins are based on 8 or fewer events, and have poor statistics.}
\end{figure}

\begin{figure}
\plotone{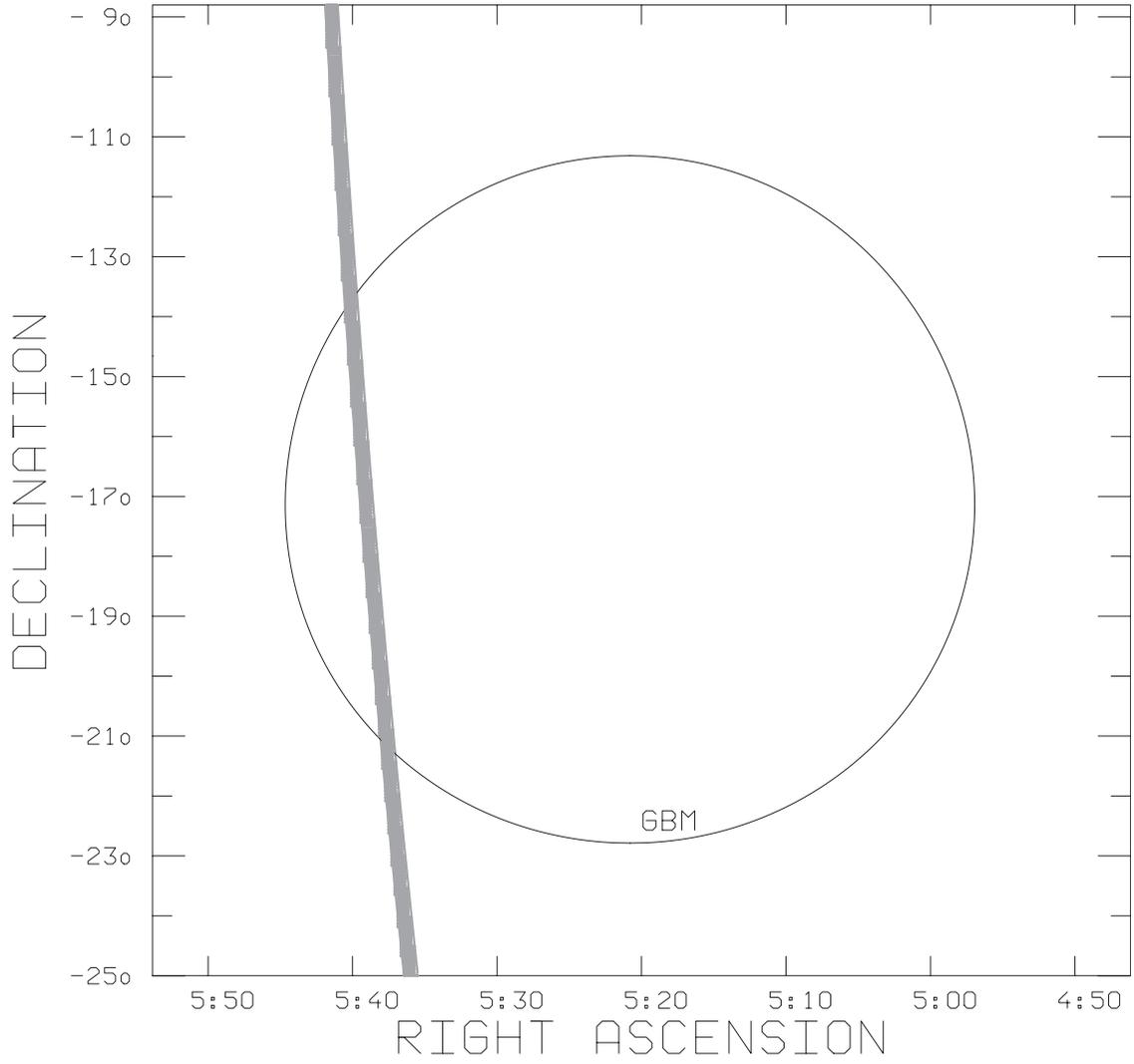}
\caption{The GBM 1$\sigma$ (statistical only) error circle for
GRB 080817B at 17:17:07 UT (GRB080817720), and the 3$\sigma$ \it Suzaku \rm-MESSENGER
triangulation annulus. 
}
\end{figure}

\begin{figure}
\plotone{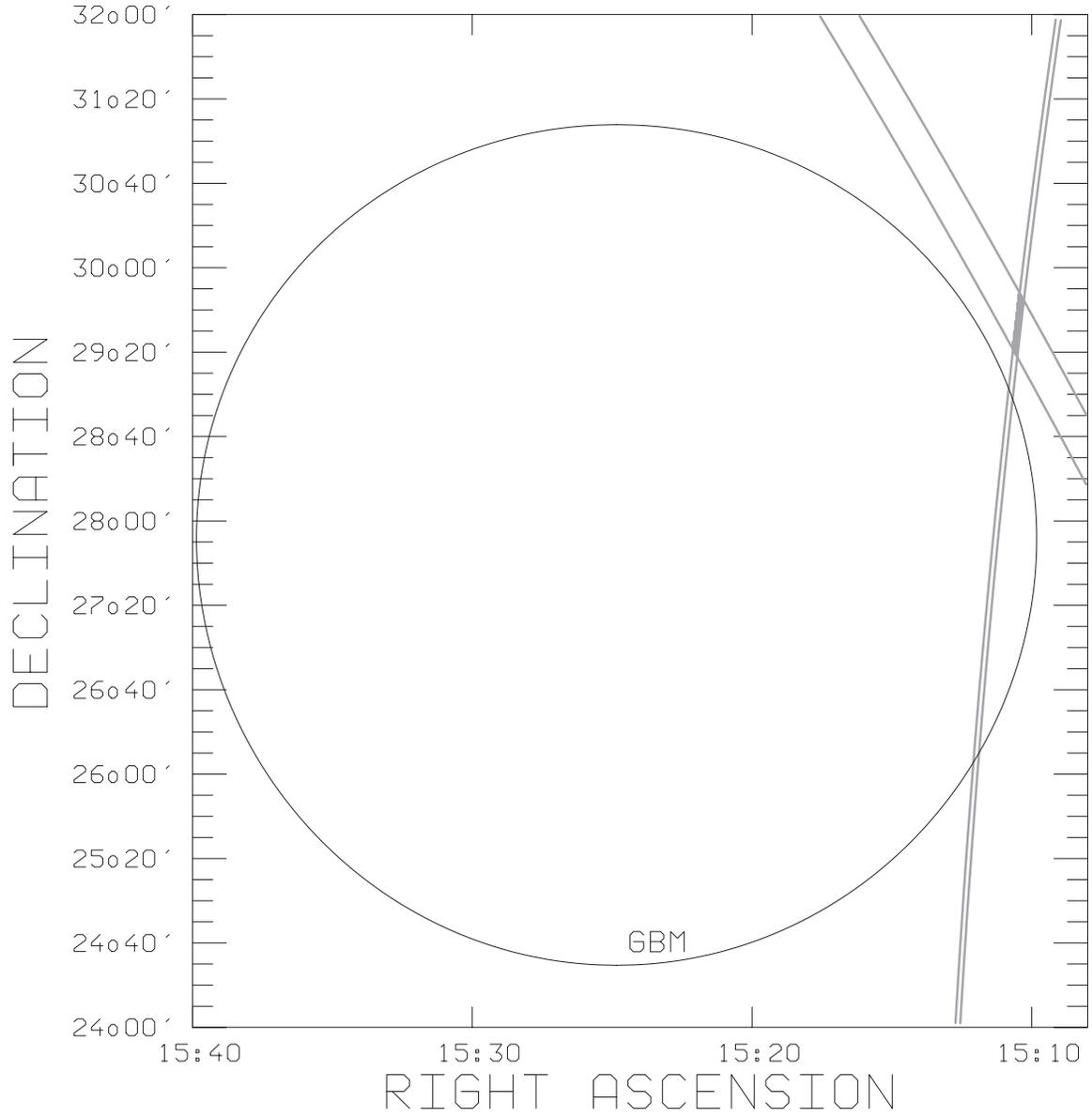}
\caption{The GBM 1$\sigma$ (statistical only) error circle for
GRB 100629 at 19:14:03 UT (GRB100629801) and the 3$\sigma$ \it Konus-Odyssey \rm and
\it Konus \rm-MESSENGER triangulation annuli.  The error box is the
shaded region. }
\end{figure}

\begin{figure}
\plotone{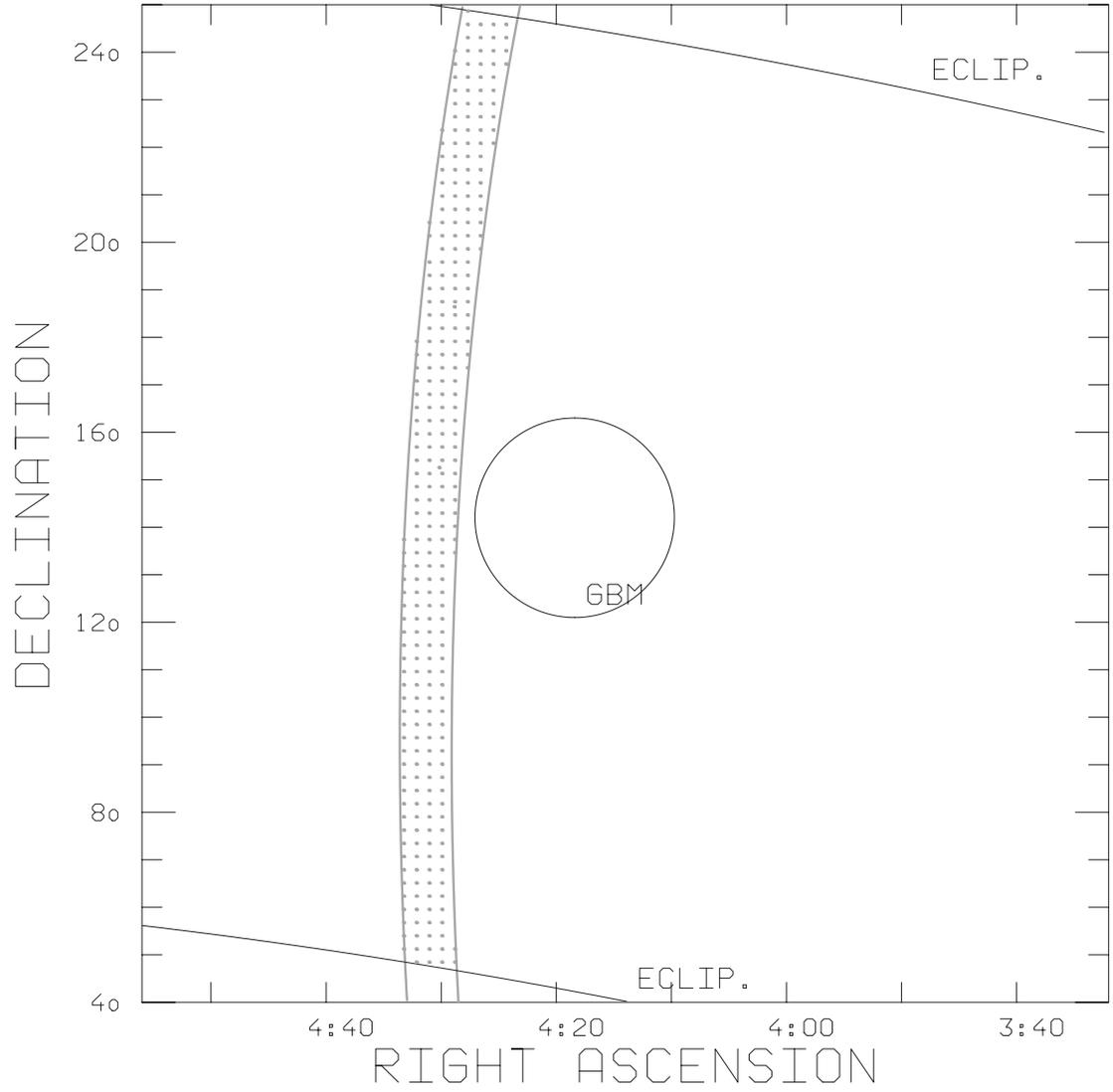}
\caption{The GBM 1$\sigma$ (statistical only) error circle for
GRB 081009B at 16:33:37 UT (GRB081009690), the 3$\sigma$ \it Konus\rm-MESSENGER
triangulation annulus, and the \it Konus \rm ecliptic latitude band. 
The error box is the shaded region.}
\end{figure}

\begin{figure}
\plotone{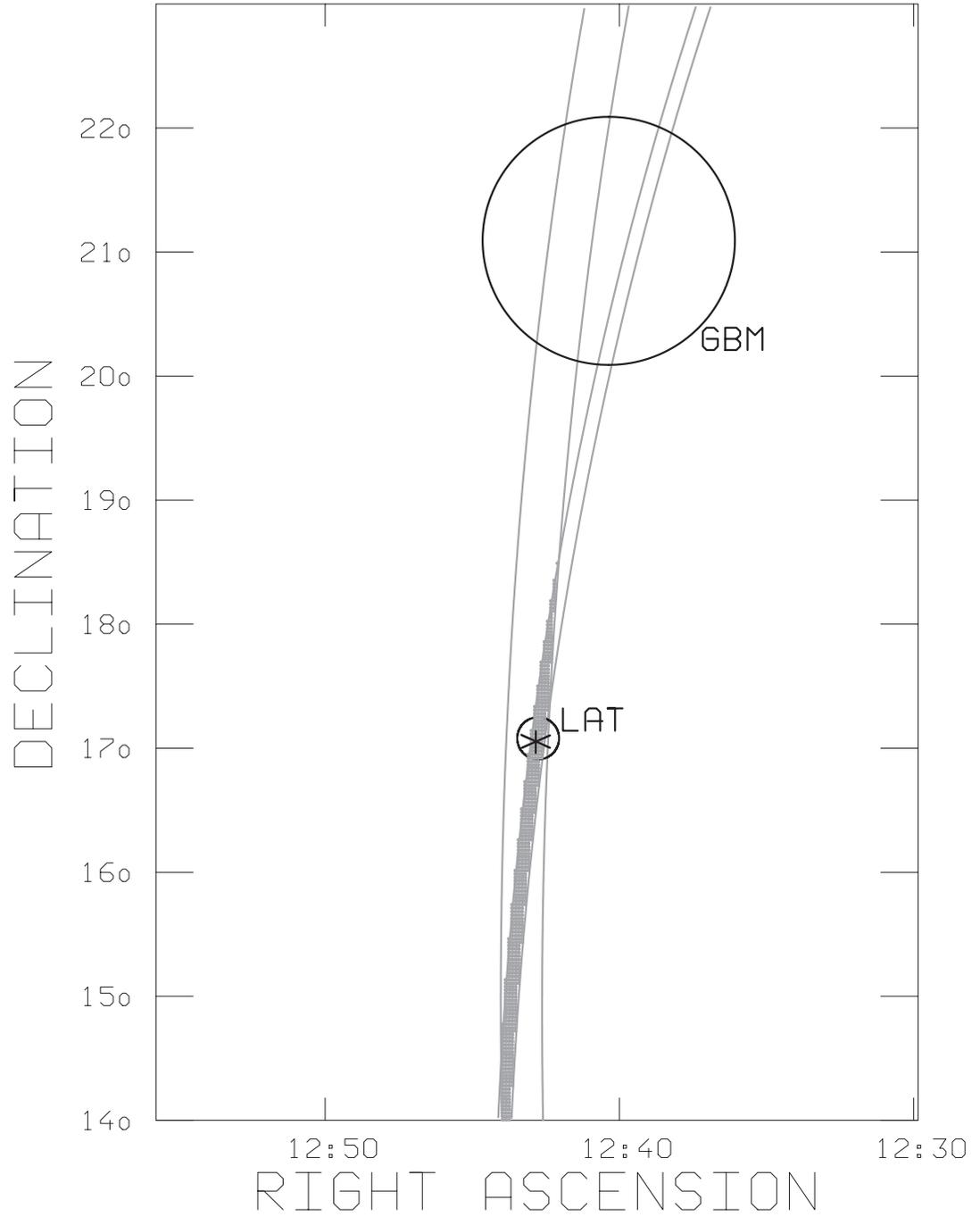}
\caption{GBM, LAT, and IPN localizations of GRB 090323 (GRB090323002).  The IPN
error box is shaded.  The position of the optical counterpart
is indicated by an asterisk.}
\end{figure}

\clearpage

\begin{deluxetable}{ccc}
\tabletypesize{\scriptsize}
\tablecaption{Recent IPN catalogs of gamma-ray bursts. \label{ipncatalogs}}
\tablewidth{0pt}
\tablehead{
\colhead{Years covered} & \colhead{Number of GRBs} & \colhead{Description}
}
\startdata

1990--1992 & 16            & \it Ulysses, Pioneer Venus Orbiter, \rm WATCH, SIGMA, PHEBUS GRBs\tablenotemark{a} \\
1990--1994 & 56            & \it Granat-\rm WATCH supplement\tablenotemark{b} \\
1991--1992 & 37            & \it Pioneer Venus Orbiter, Compton Gamma-Ray Observatory, Ulysses \rm GRBs\tablenotemark{c} \\
1991--1994 & 218           & BATSE 3B supplement\tablenotemark{d} \\
1991--2000 & 211           & BATSE untriggered burst supplement\tablenotemark{e} \\
1992--1993 & 9             & \it Mars Observer \rm GRBs\tablenotemark{f} \\
1994--1996 & 147           & BATSE 4Br supplement\tablenotemark{g} \\
1994--2012 & 271           & Konus short bursts\tablenotemark{h} \\
1996--2000 & 343           & BATSE 5B supplement\tablenotemark{i} \\
1996--2002 & 475           & \it BeppoSAX \rm supplement\tablenotemark{j} \\
2000--2006 & 226           & HETE-2 supplement\tablenotemark{k} \\
2008--2010 & 146           & GBM supplement\tablenotemark{l}
\enddata
\tablenotetext{a}{\citet{h6}}
\tablenotetext{b}{\citet{h8}}
\tablenotetext{c}{\,\citet{l2}}
\tablenotetext{d}{\citet{h3}}
\tablenotetext{e}{\citet{h9}}
\tablenotetext{f}{\,\citet{l1}}
\tablenotetext{g}{\citet{h4}}
\tablenotetext{h}{\citet{p2}}
\tablenotetext{i}{\citet{h11}}
\tablenotetext{j}{\citet{h10}}
\tablenotetext{k}{\citet{h12}}
\tablenotetext{l}{present catalog}
\end{deluxetable}

\clearpage

\clearpage

\begin{deluxetable}{cccc}
\tabletypesize{\scriptsize}
\tablecaption{IPN/GBM gamma-ray bursts. \label{detections}}
\tablewidth{0pt}
\tablehead{
\colhead{Date} & \colhead{Universal Time}\tablenotemark{a} & \colhead{GBM Identifier} \tablenotemark{b} & 
\colhead{Observed by}\tablenotemark{c} 
}
\startdata

2008 Jul 14 & 02:04:12 & GRB080714086   & Kon \\
2008 Jul 14 & 17:52:56 & GRB080714745   & AGI,INT,Kon,MES,RHE,Swi\tablenotemark{e} \\
2008 Jul 15 & 22:48:40 & GRB080715950   & AGI,Kon,MES \\
2008 Jul 17 & 13:02:35 & GRB080717543   & INT \\
2008 Jul 19 & 12:41:34 & GRB080719529   & AGI,INT,Kon \\
2008 Jul 23 & 13:22:19 & GRB080723557   & AGI,INT,Kon,MES\tablenotemark{f} \\
2008 Jul 23 & 21:56:23 & GRB080723913   & Suz \\
2008 Jul 23 & 23:37:42 & GRB080723985   & AGI,INT,Kon,MES,Suz \\
\enddata

\tablenotetext{a}{Universal time is the trigger time of a near-Earth spacecraft}
\tablenotetext{b}{Two events were not listed as GRBs in the GBM
catalog; we have confirmed however that they are valid cosmic events}
\tablenotetext{c}{
AGI: \it Astro-rivelatore Gamma a Immagini LEggero \rm (AGILE); 
INT: \it International Gamma-Ray Laboratory\rm ; 
Kon: \it Konus-Wind\rm ; 
LAT: \it Fermi Large Area Telescope\rm ;
MAXI: \it Monitor of All-sky X-ray Image\rm ;
MES: \it Mercury Surface, Space Environment, Geochemistry, and Ranging \rm mission;  
MO: \it Mars Odyssey\rm ; 
RHE: \it Ramaty High Energy Solar Spectroscopic Imager\rm ; 
RXTE: \it Rossi X-Ray Timing Explorer\rm ;
Suz: \it Suzaku\rm ; 
Swi: \it Swift\rm }
\tablenotetext{d}{Burst was outside the coded field of view of the BAT, and not localized by it}
\tablenotetext{e}{Burst was localized by Swift-BAT; IPN triangulation cannot improve on this localization}
\tablenotetext{f}{Burst was localized by SuperAGILE and INTEGRAL-ISGRI; IPN triangulation cannot improve on this localization}
\tablenotetext{g}{Burst was localized by INTEGRAL-IBIS; IPN triangulation cannot improve on this localization}
\tablenotetext{h}{Burst was localized by SuperAGILE; IPN triangulation cannot improve on this localization}

\end{deluxetable}

\begin{deluxetable}{ccccccccccccccccccccc}
\setlength{\tabcolsep}{.05in}

\rotate
\tabletypesize{\scriptsize}
\topmargin 0.5in
\tablewidth{0pt}
\tablecaption{\it IPN annuli. \label{annuli}}
\tablehead{
\multicolumn{2}{}{} & \multicolumn{3}{|c|}{GBM} &  \multicolumn{8}{|c|}{IPN} &
\multicolumn{2}{|c|}{Ecliptic} & \multicolumn{3}{|c|}{Planet} & \multicolumn{3}{|c|}{Other} \\
\colhead{GRB}&\colhead{UT}
&\colhead{$\alpha$} &\colhead{$\delta$} &\colhead{$\sigma_{stat}$}
&\colhead{$\alpha_{1}$} &\colhead{$\delta_{1}$} &\colhead{R$_{1}$} &\colhead{$\delta R_{1}$}
&\colhead{$\alpha_{2}$} &\colhead{$\delta_{2}$} &\colhead{R$_{2}$} &\colhead{$\delta R_{2}$}
&\colhead{$\beta_{1}$} &\colhead{$\beta_{2}$}
&\colhead{$\alpha$} &\colhead{$\delta$} &\colhead{R}
&\colhead{$\alpha$} &\colhead{$\delta$} &\colhead{R}
}
\startdata

\anchor{ssl.berkeley.edu/ipn3/080715/080715.eps}{080715} & 22:48:40 & 214.70 &   9.90 &  18.97 & 148.5700 &  14.4501 &  60.4476 &    .2762 & 273.7129 & -26.1484 &  71.1746 &   .2884 &  15.6 &  35.6 & ...   & ...   & ...   & ...      & ...      & ...      \\ 
\anchor{ssl.berkeley.edu/ipn3/080723B/080723B.eps}{080723} & 23:37:42 & 105.30 &  71.10 &  18.25 &  98.9192 &  24.7391 &  49.8618 &   1.5218 & 158.3648 &  10.1273 &  70.8100 &    .0376 &  40.0 &  65.0 & ...   & ...   & ...   & ...      & ...      & ...      \\ 
\anchor{ssl.berkeley.edu/ipn3/080724/080724.eps}{080724} & 09:37:40 & 358.30 &  32.90 &  18.63 &  99.4409 &  24.6844 &  84.6742 &   .6209 &  97.4219 &  21.5770 &  84.9255 &   1.1054 &  35.0 &  85.0 & 122.9 & -28.4 &  66.4 & ...      & ...      & ...      \\ 
\anchor{ssl.berkeley.edu/ipn3/080730A/080730A.eps}{080730} & 12:29:15 & 245.40 &   4.60 &  19.07 & 284.9798 & -23.2105 &  60.0101 &  11.7374 & ...      & ...      & ...      & ...      &  28.2 &  48.2 & ...   & ...   & ...   & ...      & ...      & ...      \\ 
\anchor{ssl.berkeley.edu/ipn3/080730B/080730B.eps}{080730} & 18:51:38 & 246.60 &  28.70 &  19.07 & 165.8007 &   6.4610 &  78.9123 &    .0519 & 285.2320 & -23.1364 &  67.9570 &    .8020 &  49.3 &  69.3 & ...   & ...   & ...   & ...      & ...      & ...      \\ 
\anchor{ssl.berkeley.edu/ipn3/080802/080802.eps}{080802} & 09:15:10 & 154.30 &  40.70 &  21.80 & 105.5186 &  19.2785 &  60.6310 &   1.3190 & 107.8602 &  22.3437 &  56.7793 &    .2517 &   4.6 &  90.0 & ...   & ...   & ...   & ...      & ...      & ...      \\ 
\anchor{ssl.berkeley.edu/ipn3/080803B/080803B.eps}{080803} & 18:31:20 & 300.10 &  82.80 &  25.24 & 169.8581 &   4.3476 &  84.4750 &    .4333 & 110.2713 &  17.1546 &  71.7722 &  11.7110 &  20.0 &  90.0 & ...   & ...   & ...   & ...      & ...      & ...      \\
\enddata
\end{deluxetable}

\begin{deluxetable}{cccccccccc}
\tablewidth{0pt}
\tabletypesize{\footnotesize}
\tablecaption{Number of GBM bursts observed by each IPN spacecraft. \label{ipnspacecraft}}
\tablehead{
\colhead{Konus} & \colhead{Suzaku} & \colhead{INTEGRAL} & \colhead{Swift} & \colhead{MESSENGER} & \colhead{RHESSI} &  \colhead{AGILE} &  \colhead{Odyssey} & \colhead{MAXI} & \colhead{RXTE} 
}
\startdata
         281    &           199    &           322   &             151    &          126   &                86     &            67    &              56      &           4  &           1     \\
\tableline
\enddata
\end{deluxetable}


\clearpage

\begin{thebibliography}{}
\bibitem[Aptekar et al. (1995)]{a1} Aptekar, R., Frederiks, D., Golenetskii, S., et al. 1995, \ssr, 71, 265
\bibitem[Connaughton et al. (2013)]{c1} Connaughton, V., et al. 2013, in preparation
\bibitem[Del Monte et al. (2008)]{d1} Del Monte, E., Feroci, M., Pacciani, L., et al. 2008, \aap, 478, L5
\bibitem[Gehrels et al. (2004)]{g2} Gehrels, N., Chincarini, G., Giommi, P., et al. 2004, \apj, 611, 1005
\bibitem[Gold et al. (2001)]{g1} Gold, R., Solomon, S., McNutt, R., et al. 2001, \planss, 49, 1467
\bibitem[Goldstein et al. (2012)]{g2} Goldstein, A., Burgess, J. M., Preece, R., et al. 2012, \apjs 199, 19
\bibitem[Harrison et al. (2009)]{h15} Harrison, F., Cenko, B., Frail, D., Chandra, P., and Kulkarni, S. 2009, GCN Circ. 9043
\bibitem[Hurley et al. (1999a)]{h3} Hurley, K., Briggs, M., Kippen, R. M., et al. 1999a, \apjs, 120, 399
\bibitem[Hurley et al. (1999b)]{h4} Hurley, K., Briggs, M., Kippen, R. M., et al. 1999b, \apjs, 122, 497
\bibitem[Hurley et al. (2000a)]{h6} Hurley, K., Laros, J., Brandt, S., et al. 2000a, \apjs, 533, 884
\bibitem[Hurley et al. (2000b)]{h5} Hurley, K., Kouveliotou, C., Cline, T., et al. 2000b, \apj, 537, 953
\bibitem[Hurley et al. (2000c)]{h8} Hurley, K., Lund, N., Brandt, S., et al. 2000c, \apjs, 128, 549
\bibitem[Hurley et al. (2005)]{h9} Hurley, K., Stern, B., Kommers, J., et al. 2005, \apjs, 156, 217
\bibitem[Hurley et al. (2006)]{h13} Hurley, K., Mitrofanov, I., Kozyrev, A., et al. 2006, \apjs, 164, 124
\bibitem[Hurley et al. (2009)]{h14} Hurley, K., Goldsten, J., von Kienlin, A., et al. 2009, GCN Circ. 9023
\bibitem[Hurley et al. (2010)]{h10} Hurley, K., Guidorzi, C., Frontera, F., et al. 2010, \apjs, 191, 179
\bibitem[Hurley et al. (2011a)]{h11} Hurley, K., Briggs, M., Kippen, R. M., et al. 2011, \apjs, 196, 1
\bibitem[Hurley et al. (2011b)]{h12} Hurley, K., Atteia, J.-L., Barraud, C., et al. 2011, \apjs, 197, 34
\bibitem[Kennea et al. (2009)]{k2} Kennea, J., Evans, P., and Goad, M. 2009, GCN Circ. 9024
\bibitem[Laros et al. (1997)]{l1} Laros, J., Boynton, W., Hurley, K., et al. 1997, \apjs, 110, 157
\bibitem[Laros et al. (1998)]{l2} Laros, J., Hurley, K., Fenimore, E., et al. 1998, \apjs, 118, 391 
\bibitem[Marisaldi et al. (2008)]{m1} Marisaldi, M., Labanti, C., Fuschino, F., et al. 2008, \aap, 490, 1151
\bibitem[Meegan et al. (2009)]{m2} Meegan, C., Lichti, G., Bhat, P.N., et al. 2009, \apj, 702, 791
\bibitem[Ohno et al. (2009)]{o1} Ohno, M., Cutini, S., McEnergy, J., Chiang, J., Koerding, E., and van der Horst, A. 2009, GCN Circ. 9021
\bibitem[Paciesas et al. (2012)]{p1} Paciesas, W., Meegan, C., von Kienlin, A., et al. 2012, \apjs, 199, 18
\bibitem[Pal'shin et al. (2013)]{p2} Pal'shin, V., Hurley, K., Svinkin, D., et al. 2013, arXiv:1301.3740
\bibitem[Rau et al. (2005)]{r1} Rau, A., von Kienlin, A., Hurley, K., and Lichti, G. 2005, \aap, 438, 1175
\bibitem[Smith et al. (2002)]{s1} Smith, D. M., Lin, R., Turin, P., et al. 2002, \solphys, 210, 33
\bibitem[Takahashi et al. (2007)]{t1} Takahashi, T., Abe, K., Endo, M., et al. 2007, \pasj, 59, 35
\bibitem[Tavani et al. (2009)]{t2} Tavani, M., Barbiellini, G., Argan, A., et al. 2009, \aap, 502, 995
\bibitem[Updike et al. (2009)]{u1} Updike, A., Filgas, R., Kruehler, T., Greiner, J., and McBreen, S. 2009, GCN Circ. 9026
\bibitem[Yamaoka et al. (2009)]{y1} Yamaoka, K.,  Endo, A., Enoto, T., et al. 2009, \pasj, 61, S35


\end{thebibliography}
\end{document}